\documentclass[%
 aip,
 amsmath,amssymb,
preprint,%
]{revtex4-1}

\usepackage{array}
\usepackage{multirow}
\usepackage{soul}
\usepackage{xcolor}

\usepackage{graphicx}
\usepackage{dcolumn}
\usepackage{bm}

\usepackage[utf8]{inputenc}
\usepackage[T1]{fontenc}
\usepackage{mathptmx}
\usepackage{etoolbox}
\usepackage{textcomp} 
\usepackage{gensymb} 

\makeatletter
\def\@email#1#2{%
 \endgroup
 \patchcmd{\titleblock@produce}
  {\frontmatter@RRAPformat}
  {\frontmatter@RRAPformat{\produce@RRAP{*#1\href{mailto:#2}{#2}}}\frontmatter@RRAPformat}
  {}{}
}%
\makeatother
\begin{document}


\title[Low-loss liquid metal interconnects for superconducting quantum circuits]{Low-loss liquid metal interconnects for superconducting quantum circuits}
\author{Zhancheng Yao}
\homepage{Author to whom correspondence should be addressed: yao1996@bu.edu}
\affiliation{ 
Division of Materials Science and Engineering, Boston University, Boston, Massachusetts 02215, USA
}%

\author{Martin Sandberg}%

\author{David W. Abraham}

\affiliation{%
IBM Quantum, IBM T.J. Watson Research Center, Yorktown Heights, New York 10598, USA
}

\author{David J. Bishop}

\affiliation{ 
Division of Materials Science and Engineering, Boston University, Boston, Massachusetts 02215, USA
}

\affiliation{ 
Department of Electrical and Computer Engineering, Boston University, Boston, Massachusetts 02215, USA
}%

\affiliation{ 
Department of Mechanical Engineering, Boston University, Boston, Massachusetts 02215, USA
}%

\affiliation{ 
Department of Biomedical Engineering, Boston University, Boston, Massachusetts 02215, USA
}%

\affiliation{ 
Department of Physics, Boston University, Boston, Massachusetts 02215, USA
}%

\date{\today}

\begin{abstract}
Building a modular architecture with superconducting quantum computing chips is one of the means to achieve qubit scalability, allowing the screening, selection, replacement, and integration of individual qubit modules into large quantum systems. However, the nondestructive replacement of modules within a compact architecture remains a challenge. Liquid metals, specifically gallium alloys, can be alternatives to solid-state galvanic interconnects. This is motivated by their self-healing, self-aligning, and other desirable fluidic properties, potentially enabling the nondestructive replacement of modules at room temperatures, even after operating the entire system at millikelvin regimes. In this study, we present coplanar waveguide resonators (CPWR) interconnected by gallium alloy droplets, achieving high internal quality factors up to nearly one million and demonstrating performance on par with the continuous solid-state CPWRs. Leveraging the desirable fluidic properties of gallium alloys at room temperature and their compact design, we envision a modular quantum system enabled by liquid metals.
\end{abstract}

\maketitle

Superconducting quantum circuits stand out as a leading candidate to realize practical quantum computers, as evidenced by a recent publication describing low-overhead quantum error correction protocol\cite{bravyi2024high}. As the number of physical qubits increases, a monolithic planar structure becomes impractical due to interconnect crowding within a large two-dimensional array\cite{rosenberg2017,yost2020}. To address this issue, various modular architectures integrating sub-chips are proposed\cite{gold2021entanglement,conner2021superconducting,kosen2022building,zhao2022tunable,niu2023low,yost2020,das2018cryogenic,rosenberg2017}. Furthermore, fabrication of fully yielded quantum processors could be exceedingly challenging due to individual qubit failures or frequency collisions\cite{koppinen2007complete,pop2012fabrication,hertzberg2021laser}. However, existing modular technologies are mostly long-range or permanent\cite{gold2021entanglement,yost2020,rosenberg2017,kosen2022building,zhao2022tunable,das2018cryogenic,conner2021superconducting,niu2023low}, which makes the compact architecture or nondestructive replacement of sub-chips challenging, and ultimately the failure rate of qubits will limit the scalability of the system. Therefore, technologies allowing coupling between qubits on adjacent modules and individual qubit module replacement are essential. 

Liquid metals (LM), known for their liquid state at room temperatures,  are widely studied in stretchable electronics research\cite{dickey2017stretchable}. Gallium alloys, including EGaIn (78.6\% Ga and 21.4\% In by weight) and Galistan (68.5\% Ga, 21.5\% In, and 10\% Sn by weight), are promising replacements for mercury in various applications due to their low toxicity and vapor pressure\cite{dickey2017stretchable}. They can also be alternatives to solid galvanic interconnects in superconducting quantum circuits, as their superconductivity has been studied\cite{ren2016,zhao2015single}, and their components are well-known superconductors. Combined with their self-healing\cite{li2016galinstan}, self-aligning\cite{ozutemiz2018egain}, and other desirable fluidic properties, superconducting LMs have the potential to enable nondestructive replacement of modules at room temperature, even after benchmarking the entire system at the milliKelvin regime. Meanwhile, efforts have been made to push the resolution of LM patterns to the microscale\cite{ma2023shaping,li2016galinstan,lazarus2017ultrafine,park2019high,li2015selectively,tabatabai2013liquid,kim2020nanofabrication}, making minuscule space overhead of LM interconnects possible. 

Although LM applications in microwave engineering at room temperatures have been studied\cite{wu2023circuits}, their low-temperature behavior at these frequencies remains unknown. In this work, we report the resistance of the gallium alloy (62\% Ga, 22\% In, and 16\% Sn by weight) vs. temperature above 3 K and the quality factor of such LM-bridged coplanar waveguide resonators (CPWRs) at 15 mK. Furthermore, we discuss the possible loss mechanisms induced by LMs through the preliminary data acquired from various CPWR designs. We present a step toward using LM interconnects in a quantum computer, wherein we demonstrate that such connections can be controlled, of reasonable size, and provide high-quality electrical interconnects. In future work, we will concentrate on the modularity and reuse aspects of such connections.

\begin{figure}
\includegraphics{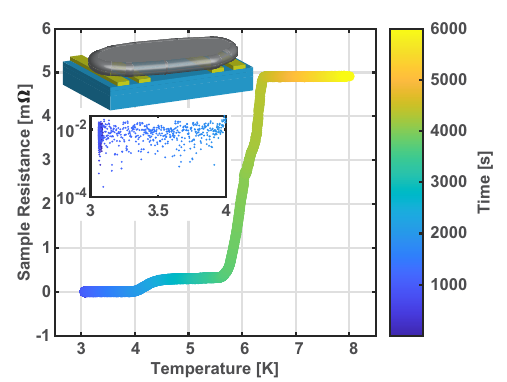}
\caption{\label{fig:sc} The four-point measurement records LM droplet (62\% Ga, 22\% In, and 16\% Sn by weight) resistance vs. temperature. The measurement starts at the base temperature of 3 K and gradually increases the temperature of the cold finger at a sufficiently low rate (1 mK/sec). The color bar indicates the time of the data point being measured in seconds. The upper inset schematic depicts a liquid metal droplet covering four gold electrodes on a high-resistivity Si wafer. The lower inset depicts a log scale resistance of the LM droplet at 3 to 4 K, showing approximately 10 \textmu $\Omega$ resistance with 7.9~\textmu $\Omega$ root mean square resistance noise of the AC resistance bridge.
}
\end{figure}

First, samples for low-temperature resistance measurement were prepared by patterning four Au electrodes on a high-resistivity Si wafer (see the supplementary material for fabrication details). An LM droplet (62\% Ga, 22\% In, and 16\% Sn by weight) was then pipetted onto the chip to cover all four gold electrodes, which are the terminals for the following four-point measurement of LM resistance.

These gold electrodes were wire-bonded to the printed circuit board (PCB) pads and connected through the cryostat feed-throughs to an AC resistance bridge with 7.9 \textmu $\Omega$ root mean square resistance noise. After reaching the cryostat's base temperature of 3 K, we increased the temperature set point by 10 mK every 10 s and recorded the resistance and temperature every 2 s. The temperature is measured at the back of the cold finger, where the front side contacts the test chip. A post-processing low-pass filter was implemented to remove the PID temperature fluctuation.

As shown in Fig.~\ref{fig:sc}, the plausible superconducting transition of the LM droplet begins at around 6.5 K and completes at about 3.8 K. Even considering its transition temperature as 3.8 K, this remains higher than that of its individual components. Although more evidence, including better resistance measurements at different external magnetic fields and specific heat characterization, is necessary to rigorously claim a superconductor, the measured response is similar to what has been reported by Ren et al.~with a similar composition of  65\% Ga, 24\% In, and 11\% Sn by weight with no external magnetic field\cite{ren2016}. Nevertheless, such a low resistance shall induce sufficiently low loss in the microwave regime, and the measured power-independent internal quality factor should give an upper bound on the resistance of the LM connections.

Next, two batches of LM-bridged CPWR samples were designed for microwave characterization. All of the CPWRs were patterned on (100)-oriented high-resistivity Si wafers with a layer of TiN (30 nm) coated Nb (200 nm) thin film. The TiN serves as a passivation layer to prevent Nb oxidation, provides a nearly oxide-free contact to LM, and acts as a diffusion barrier between Nb and LM\cite{foxen2017qubit}. The differences between the two batches of devices are highlighted in Table~\ref{tab:diff} and will be discussed later.

\begin{table}[b]
\caption{\label{tab:diff} The differences between two batches of CPWRs.  }
\begin{ruledtabular}
\begin{tabular}{ccc}
 & Batch \#1 & Batch \#2\\
\hline
LM pad width (\textmu m) & 300 & 50\\

Au/Ti pad thickness (nm) & N/A & 100/20\\

HCl treatment & No & Yes\\

Deposition method & Manual & Selective\\
\end{tabular}
\end{ruledtabular}
\end{table}

\begin{figure*}
\includegraphics[width=0.85\textwidth]{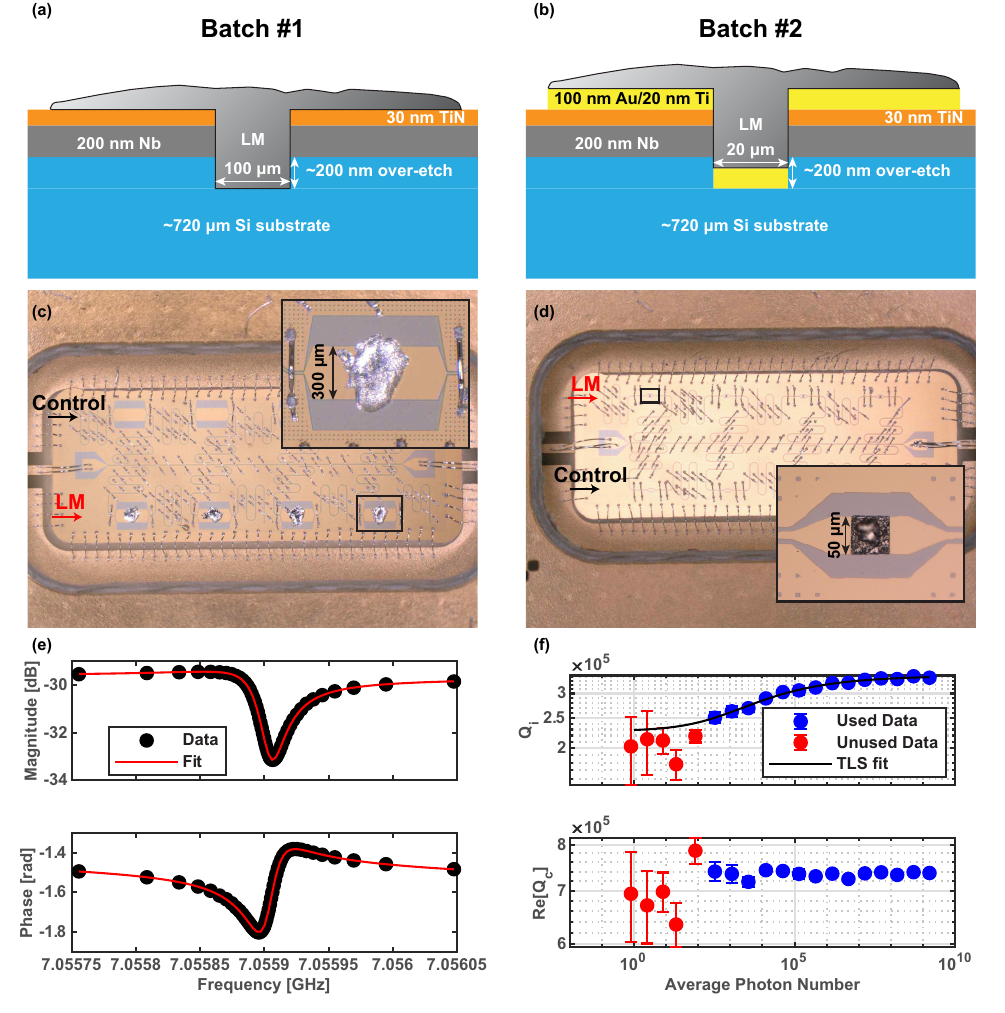}
\caption{\label{fig:rf} Microwave characterization setup and data post-processing. The material stack of the LM-bridged CPWRs pad for (a) batch \#1 with a manual LM deposition and (b) batch \#2 with a selective LM deposition. Their corresponding example photographs of wire-bonded chips are shown in (c) and (d), where the insets depict the close-ups of the LM pads. Each chip consists of four continuous CPWRs, which are controls, and four with LM droplets connecting the etched signal path gap, which are LM-bridged CPWRs. All resonators are coupled to a common TL. Numerous wire bonds are implemented to mitigate slotline modes and crosstalk. (e) The measured magnitude and phase of complex $S_{21}$ vs. the probe frequencies at 10 dBm source power, approximately corresponding to $10^9$ average photon numbers. (f) The internal quality factor, $Q_\mathrm{i}$, and the real part of the external quality factor, $Q_\mathrm{c}$, vs. the average photon number. Some fitted data may be discarded (red) depending on the confidence of the fitting results in (e). The error bars depict 95\% confidence intervals.
}
\end{figure*}

For batch \#1, the LM was manually placed to bridge the 300-by-100 \textmu m etched gap in between each half of the center strip (signal path). After the placement, another clean micropipette was then used to gently rub the LM/TiN interfaces in an attempt to break the native oxides of LM and potentially reduce the contact resistance between LM and TiN. The material stack after the manual deposition of the LM droplet is shown in Fig.~\ref{fig:rf}(a).

For batch \#2, the 50-\textmu m square LM pads were coated by a 100-nm Au with a 20-nm Ti adhesion layer after the 50-by-20~\textmu m center gap was etched. The Au pads serve as a "liquid metal-philic" material, meaning that LM tends to adhere to them more readily compared to other materials on the chip, namely selective deposition\cite{li2015selectively,kim2020nanofabrication}. A bulky LM droplet rolled back and forth on the chip until all the Au pads were covered by a layer of LM. The covered chip was then treated with a hydrochloric acid (HCl) solution bath followed by a de-ionized water rinse to remove the excess LM and oxide residue besides the Au pads. The material stack after the selective deposition of the LM droplet is shown in Fig.~\ref{fig:rf}(b).

As shown in Fig.~\ref{fig:rf}(c) and (d), each chip has four regular continuous CPWRs and four LM-bridged CPWRs coupled to a common transmission line (TL). The center strip widths and gap sizes of CPWRs and TLs are 10 and 6 \textmu m, respectively, except for the wire-bonding pads and the wide middle sections (LM pads) that accommodate the deposition of LM. All the resonators are designed as half-wavelength CPWRs, with 100-\textmu m (batch \#1) or 20-\textmu m (batch \#2) gaps for LM-bridged CPWRs at a quarter wavelength from the ends. Both short-ended and open-ended resonators were prepared to study the capacitive and inductive/resistive losses induced by LM, as their electric field and magnetic field/current are maximized at their quarter-wavelength from the ends, respectively. The continuous CPWRs serve as controls, also with wider middle sections identical to those in the LM-bridged ones. The lengths of each CPWR were designed to ensure that the resonance peaks of the control and LM-bridged CPWRs alternate across the broadband spectrum, where each pair of them is spaced by approximately 100 MHz in resonance frequency.

We utilized Ansys HFSS eigenmode to design their resonance frequencies. An analytical method\cite{besedin2018quality} was employed to estimate, followed by using the HFSS modal network to tune the coupling quality factors, $Q_\mathrm{c}$, to around $6\times10^5$. An array of 5-\textmu m vortex-pinning squares with a pitch size of 25~\textmu m was patterned on the ground conductor to reduce vortex loss. Numerous wire bonds were placed across the CPWRs and connected to the PCB ground to minimize slotline modes and crosstalk. For robust statistics, twelve chips in each batch were measured, including six with short-ended and six with open-ended CPWRs.

The devices were measured in a dilution refrigerator at 15~mK after packaging. A broadband $S_{21}$ coarse sweep was first conducted to identify individual peaks. Then, we zoomed in on individual peaks to conduct fine sweeps at source power from 10 to $-85$ dBm. We used an open-source code\cite{scresonators} to fit each trace using the diameter correction method (DCM) \cite{khalil2012} instead of the $\phi$ rotation method ($\phi$RM) \cite{gao2008}, since the latter systematically overestimates the internal quality factors, $Q_\mathrm{i}$ (see the supplementary material for a comparison of the two methods). An example of the magnitude and phase of complex $S_{21}$ vs. the probe frequency of one resonator at 10 dBm source power is shown in Fig.~\ref{fig:rf}(e).

Subsequently, we plotted the extracted $Q_\mathrm{i}$ and the real part of $Q_\mathrm{c}$ against the average photon number, $\langle n\rangle$, stored in the CPWRs as shown in Fig.~\ref{fig:rf}(f). The average photon number is calculated as\cite{bruno2015reducing,burnett2018noise}
\begin{eqnarray}
\langle n\rangle = \frac{2}{h f_0^2} \frac{Z_0}{Z_{\mathrm{r}}} \frac{Q_\mathrm{i}^2}{Q_{\mathrm{c}}} P_{\mathrm{app}},
\label{eqn:n_bar}
\end{eqnarray}
where $h$ is the Planck constant, $f_0$ is the resonance frequency of the CPWR, $Z_0$ and $Z_\mathrm{r}$ are the characteristic impedance of the TL and the CPWR, respectively, and $P_{\mathrm{app}}$ is the applied power at the TL, which is calculated from the source power given $-75$ dB attenuation distributed at the various temperature stages and cables. A total attenuation of $-80$~dB at 7~GHz is measured at room temperatures with $-70$-dB fixed attenuation and $-10$-dB temperature-dependent coaxial cable attenuation, which roughly translates to $-5$~dB at 4~K. Assuming a two-level system (TLS) loss is the only power-dependent loss mechanism, the $Q_\mathrm{i}$ is given by
\begin{eqnarray}
\frac{1}{Q_\mathrm{i}}  =  \frac{1}{Q_\mathrm{TLS}}  + \frac{1}{Q_\mathrm{other}},
\label{eqn:Qi}
\end{eqnarray}
where $Q_\mathrm{TLS}$ is the TLS quality factor, and $Q_\mathrm{other}$ is the collective quality factor induced by all the other power-independent loss channels. Then, the $Q_\mathrm{TLS}$ is given as\cite{mcrae2020materials,chiaro2016dielectric}
\begin{eqnarray}
Q_\mathrm{T L S}  =  Q_\mathrm{T L S,0} \frac{\sqrt{1+\left( \frac{\langle n\rangle}{n_\mathrm{c}} \right)^\alpha}}{\tanh \left(\frac{h f_0}{2 k_\mathrm{B} T}\right)},
\label{eqn:tls}
\end{eqnarray}
where $Q_\mathrm{T L S,0}$ is the inverse intrinsic TLS loss, $n_\mathrm{c}$ is the characteristic photon number of the TLS saturation, $\alpha$ is a coefficient accounting for a slower power dependence of the TLS saturation\cite{faoro2012internal,faoro2015interacting}, $k_\mathrm{B}$ is the Boltzmann constant, and $T$ is the temperature. An example of the fitted curve is shown in Fig.~\ref{fig:rf}(f). The DCM-fitted data with low confidence were excluded from the TLS fitting as labeled in the figure.

\begin{figure}
\includegraphics[width=0.5\textwidth]{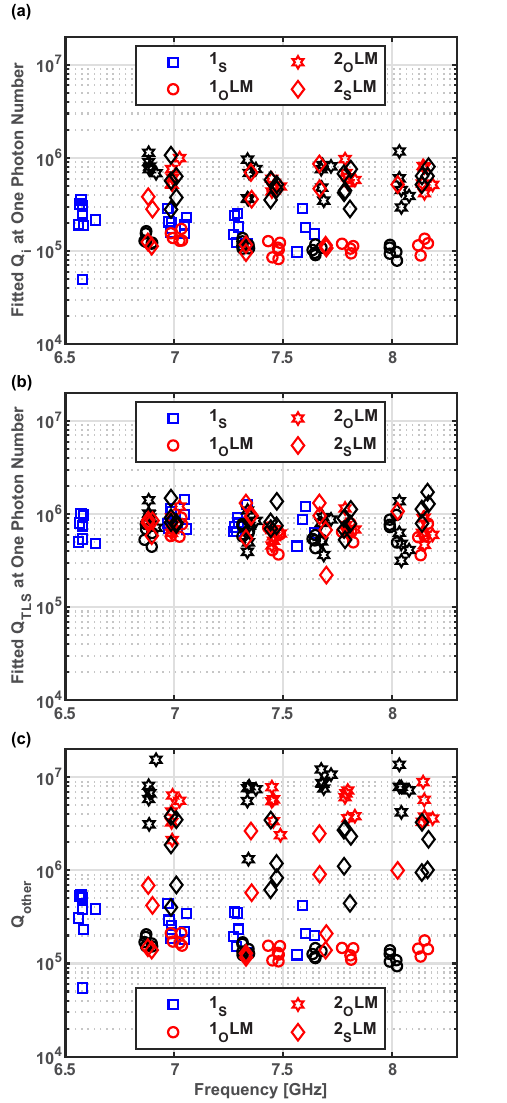}
\caption{\label{fig:sum} Fitted (a) $Q_\mathrm{i}$, (b) $Q_{\mathrm{TLS}}$, and (c) $Q_{\mathrm{other}}$ evaluated at one average photon number vs. resonance frequencies of the control (black) and LM-bridged (red) CPWRs. The numbers and subscripts in the legend denote the number of batches (1 for manual deposition and 2 for selective deposition) and the type of resonators (S for short-ended and O for open-ended), respectively. The control and LM-bridged resonator data are indistinguishable from the short-ended resonators in batch \#1, so we instead use blue color to represent data from both control and LM-bridged resonators.
}
\end{figure}

\begin{figure}
\includegraphics{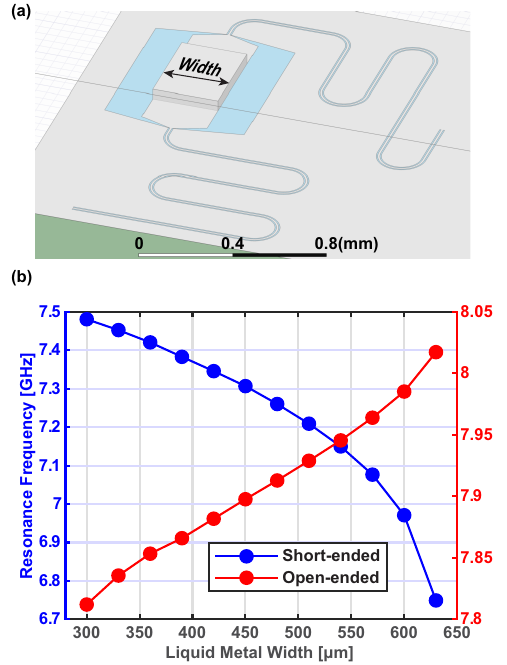}
\caption{\label{fig:sft} HFSS simulation of the width dependence of resonance frequencies of the LM-bridged CPWRs. (a) The HFSS Eigenmode model of an open-ended LM-bridged CPWR, where the LM droplet is simplified as a rectangular cuboid perfect conductor and located at the center along the signal path. The width of the LM droplet varies in the direction normal to the signal path and parallel to the substrate plane. (b) The resonance frequency vs. the width of LM metal, which increases from the signal path width (300 \textmu m) for short-ended (blue) and open-ended (red) resonators.
}
\end{figure}

The fitted Eq.~(\ref{eqn:Qi}) was evaluated at one average photon number, and the $Q_\mathrm{i}$, $Q_\mathrm{TLS}$, and $Q_\mathrm{other}$ were plotted against the $f_0$ as shown in Fig.~\ref{fig:sum}(a), (b), and (c), respectively. Each marker shape denotes a type of CPWR (batch number and open/short), and the color denotes whether it is from control (black) or LM (red). The data markers denoting short-ended resonators in batch \#1, however, are all blue due to the challenge of distinguishing resonance peaks of control and LM-bridged CPWRs. We argue that this is from the artificial frequency shifts induced by larger LM droplet widths as shown in Fig.~\ref{fig:rf}(c) inset. We simulated the width vs. the resonance frequency with a model as shown in Fig.~\ref{fig:sft}(a). Figure~\ref{fig:sft}(b) shows that for short-ended resonators, the resonance frequency decreases when the LM width increases. For the specific type of devices, the resonance frequencies of LM-bridged CPWRs are only 100 MHz larger than their paired controls. Therefore, it is likely that the designed 100 MHz gap can vanish or even be reversed if the actual width of the manually deposited LM is more than 100 \textmu m larger than the designed value. We later swapped the resonance peaks of control and LM-bridged resonators for the short-ended ones in batch \#2 so that such a frequency decrease in the LM-bridged resonators would only result in a gap larger than its designed value.

\renewcommand{\arraystretch}{1.5}
\begin{table}
\caption{\label{tab:sum} The average values of fitted quality factors at one average photon number in Fig.~\ref{fig:sum}. CT and LM refer to control and LM-bridged CPWRs, respectively. LM-induced loss from short-ended resonators is mostly capacitive, while its loss from open-ended resonators is mostly resistive/inductive. The uncertainties reflect the sample-to-sample variation.}
\begin{tabular}{|c|c|c|c|c|c|}
\hline
 & & & $Q_\mathrm{i}$ ($10^5$) & $Q_\mathrm{TLS}$ ($10^5$) & $Q_\mathrm{other}$ ($10^5$)                    \\ \hline
\parbox[t]{5mm}{\multirow{4}{*}{\rotatebox[origin=c]{90}{Batch \#1}}}   & \multicolumn{1}{c|}{\multirow{2}{*}{Short}} & \multirow{2}{*}{\textcolor{blue}{N/A}} & \multirow{2}{*}{\textcolor{blue}{$2.02\pm 0.71$}} & \multirow{2}{*}{\textcolor{blue}{$8.35 \pm 2.52$}} & \multirow{2}{*}{\textcolor{blue}{$2.87 \pm 1.31$}} \\ 
                             &                                               &             &            &           & \\ \cline{2-6} 
                             & \multirow{2}{*}{Open}                        & \textcolor{red}{LM} & \textcolor{red}{$1.18 \pm 0.23$} & \textcolor{red}{$6.22 \pm 1.47$} & \textcolor{red}{$1.48 \pm 0.31$} \\ \cline{3-6} 
                             &                                               & CT & $1.15 \pm 0.20$ & $6.58 \pm 1.44$ & $1.41 \pm 0.28$ \\ \hline
\parbox[t]{5mm}{\multirow{4}{*}{\rotatebox[origin=c]{90}{Batch \#2}}}   & \multirow{2}{*}{Short}                       & \textcolor{red}{LM} & \textcolor{red}{$3.26 \pm 2.52$} & \textcolor{red}{$8.56 \pm 3.03$} & \textcolor{red}{$7.39 \pm 8.67$} \\ \cline{3-6} 
                             &                                               & CT & $5.57 \pm 1.98$ & $9.67 \pm 3.18$ & $18.43 \pm 11.79$ \\ \cline{2-6} 
                             & \multirow{2}{*}{Open}                       & \textcolor{red}{LM} & \textcolor{red}{$6.17 \pm 1.67$} & \textcolor{red}{$7.19 \pm 2.04$} & \textcolor{red}{$49.67 \pm 18.71$} \\ \cline{3-6} 
                             &                                               & CT & $6.78 \pm 2.61$ & $7.67 \pm 3.20$ & $78.78 \pm 33.07$ \\ \hline
\end{tabular}
\end{table}

As shown in Table~\ref{tab:sum}, average $Q_\mathrm{i}$ values of $3.26\times10^5$ and $6.17\times10^5$ are achieved for batch \#2 short- and open-ended LM-bridged CPWRs, respectively. These values surpass the $Q_\mathrm{1}$ of TSV-interrupted resonators ($1-3\times10^5$) reported by Mallek et al.\cite{mallek2021fabrication} The best-performing LM-bridged CPWR, yielding nearly one million $Q_\mathrm{i}$, as shown in Fig.~\ref{fig:sum}(a), exceeds the $Q_\mathrm{i}$ of Al coaxial cables ($8.1\times10^5$) reported by Niu et al., despite comparing the 11th mode instead of the fundamental mode of the long cable\cite{niu2023low}. Furthermore, a direct comparison with indium bumps reported by Rosenberg et al. can be made by calculating the reduced $Q_\mathrm{i}$, $Q_\mathrm{i,reduced} = 1/(1/Q_\mathrm{i,LM}-1/Q_\mathrm{i,CT})$\cite{rosenberg2017}. The $Q_\mathrm{i,reduced}$ of $7.86\times10^5$ and $6.86\times10^6$ for batch \#2 short- and open-ended LM-bridged resonators, respectively, are comparable to the reported values of indium bumps ($0.8-8\times10^5$)\cite{rosenberg2017}. However, even the highest-quality control resonators do not surpass the one-million-Q threshold as significantly as state-of-the-art resonators, due to the limitations in the processing and material choices\cite{crowley2023disentangling}. However, a noticeable improvement in $Q_\mathrm{i}$ is observed from batches \#1 to \#2. Specifically, orders of magnitude improvement in $Q_\mathrm{other}$ is observed while $Q_\mathrm{TLS}$ remains about the same. As summarized in Table~\ref{tab:diff}, the major differences from batches \#1 to \#2 are as follows:
\begin{itemize}
\item The reduction in LM pad width from 300 to 50 \textmu m, which should affect size-dependent loss.
\item The HCl treatment, which could remove TLS sources like oxides that are reactive to HCl. In principle, only LM and its oxides can be affected by HCl.
\item The gold pad, specifically for LM-bridged CPWRs, which could introduce a reduction in $Q_\mathrm{i}$.
\end{itemize}
The gold pads may induce different types of loss depending on whether the actual current is through them or runs parallel to them. A future study is necessary to inspect the actual structure and further determine the detailed mechanisms. Given the known changes in the process, further breakdown of the loss mechanism may be discussed as follows.

As the overall $Q_\mathrm{other}$ is significantly improved from batches \#1 to \#2, the possible gold pad resistive loss seems to be a minor factor. Furthermore, size reduction could increase the TLS surface participation ratio by having a greater fraction of the electric field energy coupled to interface defects, thus increasing TLS loss\cite{gao2008experimental,crowley2023disentangling}. However, no decrease in $Q_\mathrm{TLS}$ from batches \#1 to \#2 has been observed, indicating that either the size reduction and HCl treatment both had little influence on TLS loss, or the two competing factors canceled each other out. This remains poorly understood without further verification.

If HCl treatment makes a negligible impact on $Q_\mathrm{other}$, the most likely improvement from batches \#1 to \#2 is from a reduction in vortex and radiative loss, as both should decrease with a size reduction. Wider center conductors are more susceptible to vortex trapping\cite{stan2004critical,mcrae2020materials}, and both larger conductor and gap sizes can yield more radiation\cite{sage2011study,mcrae2020materials}. Here, vortex loss refers to dissipation from the trapped magnetic flux, which is mostly inductive, while radiation can be propagated from fluctuating electric and magnetic fields, which can be both capacitive and inductive. The improvement in $Q_\mathrm{other}$ for open-ended resonators due to the LM pad size reduction is larger compared to short-ended resonators. Therefore, the inductive contribution, consisting of vortex and radiative factors, of such an improvement is likely larger. Additional geometry designs are required to separate the two factors\cite{sage2011study}.

For open-ended resonators in both batches, the introduction of LM has yet to compromise the $Q_\mathrm{i}$ significantly. Although the $Q_\mathrm{other}$ in batch \#2 decreased upon LM deposition, it is less dominant compared to the TLS loss. This drop may consist of both LM and gold contribution but requires further investigation. Conversely, for short-ended resonators in batch \#2, the major reduction in $Q_\mathrm{i}$ comes from the decreased $Q_\mathrm{other}$, implying an LM-induced power-independent capacitive contribution. Such a loss could be radiative loss or due to parasitic modes\cite{mcrae2020materials}. Considering the large sample-to-sample variation, this could also be explained by a nonideal fabrication yield.

In summary, the low-loss LM interconnects showcase the building block of a modular quantum system. The comparable performance of continuous and LM-bridged resonators unlocks the possibilities of LM-bridged bus resonators, control, and readout lines across modules. With LM bumps serving as quantum interconnects across chips, nondestructive replacement of individual modules might be possible at room temperatures, even after low-temperature benchmarking of the entire system.

Future work should aim to demonstrate a true LM quantum connection between two chips. To achieve this, a more controllable fabrication process should be developed and implemented. Smaller droplet sizes, more precise placement and alignment, and better surface control of the liquid metal can improve space utilization and repeatability, facilitating understanding and reducing its TLS loss. Additionally, introducing more variables, including temperature\cite{crowley2023disentangling} and geometry\cite{stan2004critical,crowley2023disentangling,sage2011study}, would assist researchers in disentangling the loss mechanisms induced by liquid metal, thus providing guidance on solutions to improve its quality factor. Finally, increasing the quality factor of the non-LM part of the resonators, either through refined processing or other material choices, could further improve the characterization of the LM loss contribution.

\setcounter{figure}{0}
\renewcommand{\figurename}{Fig.}
\renewcommand{\thefigure}{S\arabic{figure}}

\section*{Supplementary Material}

The supplementary material describes more detailed fabrication methods, a comparison of DCM and $\phi$RM, and more examples of the TLS fit.

\renewcommand{\figurename}{Fig.}
\renewcommand{\thefigure}{S\arabic{figure}}

\subsection{Fabrication}

\subsubsection{\label{sec:scfab}The liquid metal resistance measurement chip fabrication}

An 8-inch, (100)-oriented, high-resistivity, single-side polished silicon wafer is first cleaved into four quarters. Then, one quarter was taken and cleaned successively with Acetone, isopropanol (IPA), and de-ionized (DI) water. Following the solvent cleaning, the wafer was mounted on a spin coater and spin-coated a layer of lift-off photoresist (LOR 3A). It is then soft-baked for 4 minutes at 180 \degree C. On top of the baked LOR resist, another layer of positive photoresist (Shipley S1805) was spin-coated and soft-baked for 1 minute at 115 \degree C. The bi-layer photoresist creates undercuts of LOR resist during the development process, thus allowing lift-off of the metal film deposited in the next step.

The pattern of gold electrodes is defined by laser direct writing (Heidelberg MLA 150) after soft-bake. Next, we developed the exposed resist using a metal-ion-free developer (Shipley CD-26). After the DI water rinse and nitrogen drying, the patterned wafer was mounted onto the sample holder and installed in an electron-beam (E-beam) evaporator (CHA Solution). A 20-nm layer of Ti and a 200-nm layer of Au were then deposited at high vacuum as an adhesion layer and electrodes, respectively.

Following the deposition, the wafer was immersed in an organic compound solution (Microposit Remover 1165) at 95~\degree C until the photoresist was fully dissolved in the solution bath. Finally, the wafer was coated by another protective layer of photoresist and then diced into individual chips, with each containing four gold electrodes. The diced chips were immersed in Remover 1165 at 95 \degree C again to dissolve the protective photoresist.

\subsubsection{\label{sec:batch1}Batch \#1 chips fabrication}

An 8-inch, (100)-oriented, high-resistivity, single-side polished silicon wafer was Huang cleaned, and a layer of 200~nm Nb was sputtered onto the wafer. Another layer of 30 nm TiN was also sputtered onto the Nb thin film without breaking the vacuum. The TiN serves as a passivation layer to prevent Nb oxidation, provides a nearly oxide-free contact to LM, and acts as a diffusion barrier between Nb and LM\cite{foxen2017qubit}.

The metalized wafer was coated with a positive photoresist (Shipley S1818), exposed by the direct laser writing, and then developed by a Shipley MF-319 developer. The gaps of coplanar waveguide resonators (CPWRs), the common transmission line (TL), and vortex trapping holes were defined by the aforementioned lithography process. Further, we used a reactive ion etcher (Plasma-Therm 790) with a mixture of SF\textsubscript{6}/Ar to etch through the exposed TiN and Nb. We over-etched approximately 200 nm of the silicon substrate to reduce the two-level system (TLS) loss contribution from the substrate-air interfaces\cite{bruno2015reducing,sandberg2012etch}. The individual chips were diced from the patterned wafer with a photoresist protective layer, followed by the Remover 1165 bath to remove the resist.

\subsubsection{\label{sec:batch2}Batch \#2 chips fabrication}

A similar fabrication process was implemented before the dicing process. Next, the Au LM pads are defined by the bi-layer resists and the direct writing. A layer of 20-nm Ti followed by 100-nm Au was E-beam evaporated under a high vacuum. Finally, a similar lift-off and dicing process were applied.

\subsection{Comparison of DCM and $\phi$RM}

\begin{figure*}
	\includegraphics{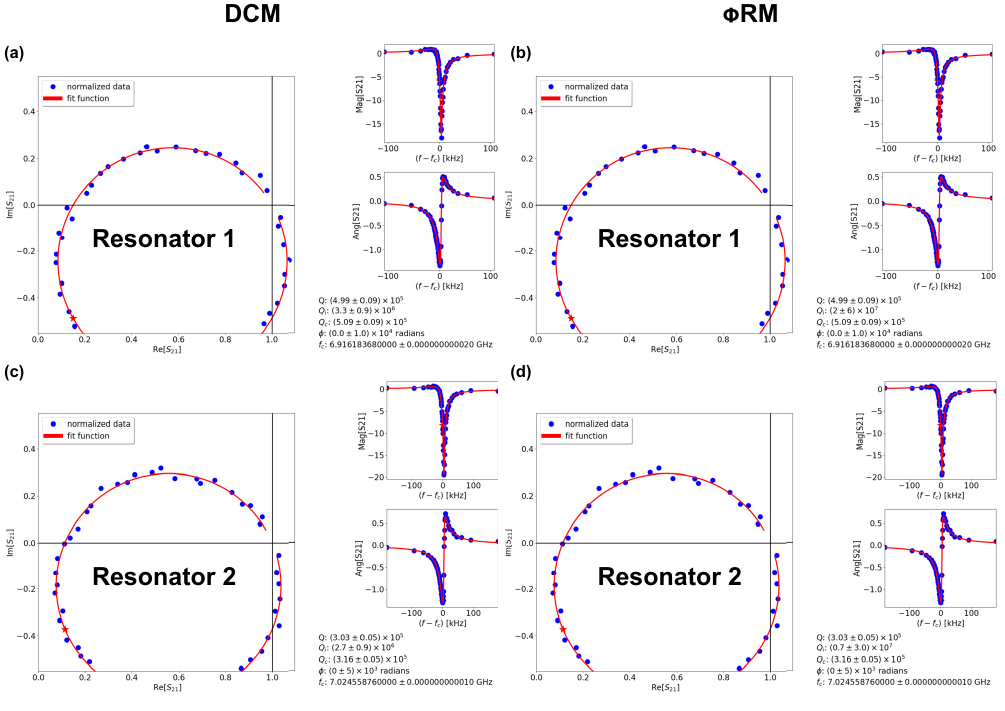}
	\caption{\label{fig:comp} Comparison of fitted $Q_\mathrm{i}$ using DCM and $\phi$RM. (a) DCM-fitted and (b) $\phi$RM-fitted $Q_\mathrm{i}$ of Resonator 1 are $(3.3 \pm 0.9)\times10^6$ and $(2 \pm 6)\times10^7$, respectively. (c) DCM-fitted and (d) $\phi$RM-fitted $Q_\mathrm{i}$ of Resonator 2 are $(2.7 \pm 0.9)\times10^6$ and $(0.7 \pm 3)\times10^7$, respectively. In each case, $\phi$RM overestimates $Q_\mathrm{i}$ significantly compared to DCM.
	}
\end{figure*}

In our data set, $\phi$RM systematically overestimates $Q_\mathrm{i}$ to different extents compared to DCM. Two examples are shown in Fig.~\ref{fig:comp}, where the raw data at the top and bottom panels are from Resonator 1 and 2 at $-30$ dBm source power, respectively. The left and right side fit parameters are generated by DCM and $\phi$RM using the open-source code \cite{scresonators}, respectively. As shown in the figure, DCM yielded more realistic $Q_\mathrm{i}$ magnitudes and smaller confidence intervals than the $\phi$RM ones. A more thorough comparison of the two methods has been discussed by Khalil~et~al. \cite{khalil2012}

\subsection{Examples of TLS fit}

\begin{figure*}
	\includegraphics{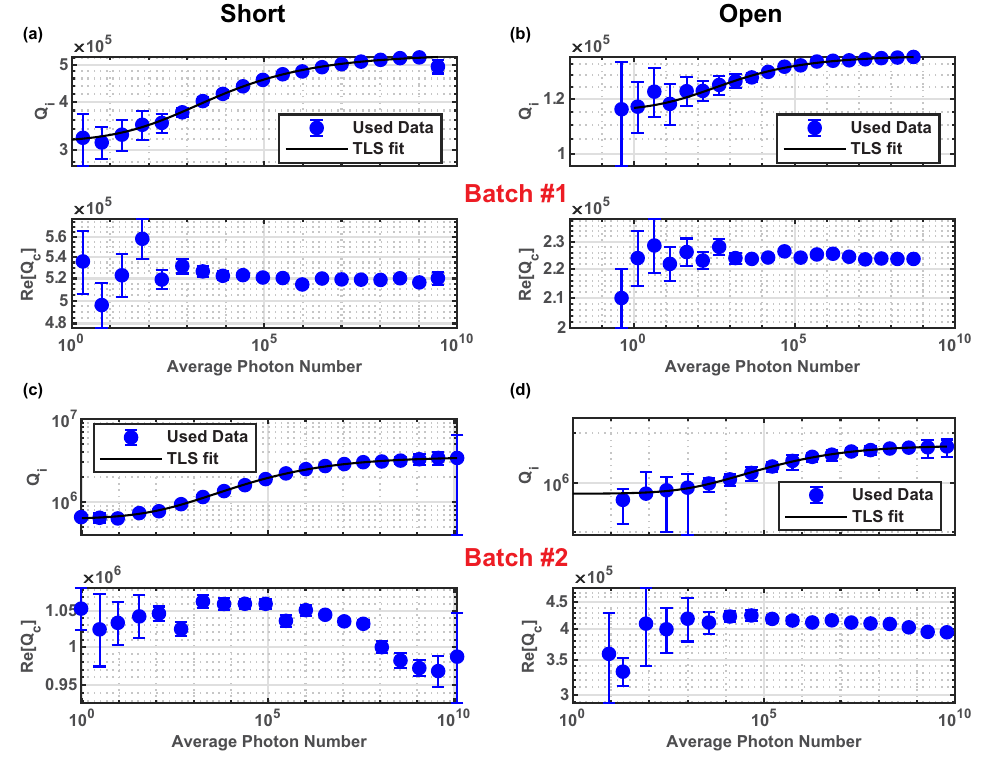}
	\caption{\label{fig:exTLS} Examples of TLS fit from (a) a short-ended resonator from Batch \#1, (b) an open-ended resonator from Batch \#1, (c) a short-ended resonator from Batch \#2, and (d) an open-ended resonator from Batch \#2. The error bars depict 95 \% confidence intervals.
	}
\end{figure*}

Here, we provide more examples of TLS fit as shown in Fig.~\ref{fig:exTLS}.

\setlength{\parskip}{10pt}

This work was supported by an IBM-sponsored research award No.~W2178130.~The authors acknowledge the use of facilities at the Boston University Photonics Center and the Harvard University Center for Nanoscale Systems (CNS). CNS is a member of the National Nanotechnology Coordinated Infrastructure Network (NNCI), which is supported by the National Science Foundation under NSF award No.~ECCS-2025158.

\setlength{\parskip}{0pt}

\section*{Data Availability}

The data that support the findings of this study are available from the corresponding author upon reasonable request.

\nocite{*}
\bibliography{aipsamp}

\end{document}